\documentclass[%
 reprint,
 amsmath,amssymb,
 aps,
floatfix
]{revtex4-2}
\usepackage[autostyle]{csquotes}
\usepackage{graphicx}
\usepackage{dcolumn}
\usepackage{bm}
\usepackage{hyperref}
\usepackage{gensymb}
\usepackage{amsmath}
\hypersetup{colorlinks = True,
linkcolor = blue,
filecolor = magenta,
urlcolor = blue,
citecolor = blue}

\begin{document}

\preprint{APS/123-QED}

\title{Tracking phase entanglement during propagation of downconverted photons}

\author{Rounak Chatterjee}
    \author{Vikas S Bhat}
    \author{Kiran Bajar}
    \author{Sushil Mujumdar}
     \email{mujumdar@tifr.res.in; http://www.tifr.res.in/~nomol}
    \affiliation{%
     Tata Institute of Fundamental Research, 400005 Mumbai, India
    }%

\date{\today}

\begin{abstract}
High-dimensional entanglement, in the form of transverse spatial correlation between a pair of photons generated via spontaneous parametric downconversion, is not only a valuable resource in many academic and real-life applications, but also provides access to several intriguing quantum phenomena. One such non-intuitive phenomenon is phase entanglement, in which the biphoton state is correlated in the complex phase of its wavefunction. This state, which emerges during the propagation of the biphoton wavefunction, exhibits no position correlation, yet retains full entanglement. In this work, we experimentally explore this state in two distinct ways. The first is by tracking the vanishing spatial photon number correlation over propagation distances lying in $\left[0,\infty\right)$, folded into a finite range using single-lens imaging. These observations show excellent agreement with our theoretical predictions based on the Double Gaussian (DG) approximation of the biphoton state. The second approach involves performing a two-photon interference experiment using a double slit and this state, which reveals the correlated phase front. We show, both theoretically and experimentally, that the observed two-photon interference structure is markedly different from that produced by position-correlated photons, as confirmed by computing the joint probability distribution of photons (JPD) and related metrics. Such interference using phase-entangled light has implications that open avenues for advanced experiments and applications in the field of spatial entanglement.
\end{abstract}
\maketitle

\section{Introduction}
\label{sec:Intro}

The phenomenon of spontaneous parametric down-conversion (SPDC) generates many quantum states that are not only academically intriguing but also rich in real-world applications. Examples include heralded single-photon sources \cite{Heralded2004} and polarization-entangled sources \cite{Kwiat1995}, which are used in fields such as quantum communication \cite{Gisin2002}. SPDC is also employed in foundational experiments, like the Hong-Ou-Mandel effect \cite{HOM1987} and quantum teleportation \cite{Bouwmeester1997}, which serve as cornerstones for numerous applications in quantum information processing and potential commercial technologies like quantum repeaters \cite{Azuma2023}. Additionally, SPDC is known to create bipartite entanglement in higher discrete dimensions, such as orbital angular momentum \cite{Mair2001,Krenn2017}, and in continuous dimensions, such as spatial entanglement, manifesting as correlations in the position and momentum of photon pairs \cite{Howell2004, Walborn2010,howell2016}. The spatial entanglement property offers a wide variety of quantum states that can be tailored to specific applications such as light propagation through disordered media \cite{Peeters2010,Lib2022,Devaux2023,Courme2023}, quantum imaging \cite{Abouraddy2001,Santos2003,Lemos2014,Gatti2008,Soro2021}, and recent developments in quantum microscopy \cite{He2023,Karimi2024}.
\par In most applications of spatial entanglement, either position correlations or momentum anti-correlations of photons have been employed \cite{Howell2004} to extract properties of the medium or object under study. However, spatial entanglement also gives rise to a non-intuitive yet potentially useful quantum state known as the phase-entangled state \cite{Chan2007}. This state is unique in that the biphoton wavefunction exhibits no intensity correlations, i.e, no position, yet the entanglement is preserved since the phase of the wavefunction is non-separable. Intuitively, one can imagine this two-photon state to have a combined  well-defined phase front, while individual photon phase-front is completely randomized. Another outlook to the physical implication of this state can be found in  \cite{Chan2004}, where it was theoretically demonstrated that the phase entanglement manifests as a correlation between the position of one photon and the momentum of the other. Although a general phase-entangled state is difficult to associate with any direct physical observable, its mathematical implications are profound. Any physical phenomena reliant on interferometric measurements, or reliant on the critical role of the phase, could benefit from a deeper exploration of this quantum state. Notably, this phase-entangled state naturally arises as the SPDC wavefunction propagates through space and can be accessed using appropriate imaging systems—a concept we aim to explore in this work.

\par However, characterizing these states is challenging and has not yet been experimentally demonstrated. To date, they have either been simulated theoretically \cite{Just2013}, or measurements at some distance from the near and far fields \cite{Reichert2017} have been performed, but no direct attempts to probe the intermediate phase entanglement plane have been undertaken. In this work, we propose two theoretically justified methods to probe the phase entanglement plane. The first is by examining the absence of intensity correlation at this plane. This approach involves propagating the initial position-correlated biphoton state—justifiably approximated as a double Gaussian, and measuring the subsequent loss of intensity correlation as it approaches a critical distance. Intuitively, by the argument of preservation of entanglement during propagation, we can justify that, at this critical distance, all the correlation must be present in the quantum phase. We theoretically justify this proposition in sec.\ref{sec:theory} as well as experimentally verify it in sec.\ref{sec:Exp}. The propagation of the biphoton wavefunction over the full range from near to far field is tracked by folding the propagation distance using single-lens imaging. This approach not only lets us probe the state but also gives us a way to deterministically create it with desirable parameters.

\par The second approach to showcase phase entanglement is by performing a two-photon interference experiment using a simple double-slit with this state. It is well known that pure two-photon interference effects shown by near- and far-field biphoton states are solely mediated by inherent amplitude correlations \cite{HOM1987,Monken1999,Abouraddy2001_DS,Dixon2010}. However, in our case, we show theoretically as well as experimentally that for phase-entangled light, the pure two-photon interference effect is entirely mediated by the inherent phase correlations, since a phase-entangled state doesn't have any amplitude correlations to begin with. In sec.\ref{sec:DS setup and theory}, we mathematically formalize this concept using the formalism described in \cite{Saleh2000} and also verify our conclusions experimentally. Formal study of any spatial quantum state is usually performed using interference effects, and this interference effect arising from inherent quantum phase correlations has not previously been observed in experiments. We believe this insight opens the door to advanced experimental studies and potential applications of such states in various fields such as the study of disordered media and imaging with phase objects.

\section{Phase Entangled Double Gaussian }
\label{sec:theory}
SPDC from a specially cut non-linear crystal of length $L$, pumped by a collimated laser (wavelength = $\lambda_p$, refractive index in the crystal = $n_p$, field profile $E_p(\textbf{r})$), creates near-degenerate photon pairs at twice the wavelength. These photon pairs are spatially entangled in their position and momentum variables and propagate co-linearly with the pump beam. The wavefunction in the momentum variables of the two photons ($\textbf{p}_1, \textbf{p}_2$) can be written as\cite{Walborn2003}:
\begin{equation}
\label{eqn:SPDC original}
    \psi_k(\textbf{p}_1,\textbf{p}_2) = \widetilde{E}_p(\textbf{p}_1+\textbf{p}_2) \text{ sinc}\left(\frac{L\lambda_p}{8\pi n_p}|\textbf{p}_1-\textbf{p}_2|^2\right)
\end{equation}
Here, $\widetilde{E}_p$ represents the Fourier transform $(\mathcal{F})$ of $E_p$. The wavefunction in position variables is given by $\psi(\textbf{r}_1, \textbf{r}_2) = \mathcal{F}\left\{\psi_k(\textbf{p}_1, \textbf{p}_2)\right\}$. Assuming the pump beam is an azimuthally symmetric Gaussian beam with a waist $\sigma_p$, and that the waist is located exactly at the center of the crystal, we can approximate the wavefunction as a DG in either the position or momentum variables \cite{law2004,Monken1998}. This coordinate system is advantageous since all planes in the crystal contribute coherently to the central plane\cite{Pires2009}). The 1D position wavefunction is then given as:

\begin{eqnarray}
\label{eqn:DG}
 &\psi(x_1,x_2) =  \frac{1}{\sqrt{\pi \sigma_-\sigma_+}}\exp{\left(-\frac{(x_1-x_2)^2}{4\sigma_-^2}-\frac{(x_1+x_2)^2}{4\sigma_+^2}\right)}
\end{eqnarray}
where $\sigma_- = \sqrt{\frac{L\lambda_p}{6\pi n_p}}$ and $\sigma_+ = \frac{\sigma_p}{2}$. This form possesses several features that are both theoretically tractable and experimentally measurable. For example, the widths $\sigma_{\pm}$ provide information about the degree of position correlation, momentum anti-correlation\cite{Edgar2012,Moreau2012}, and the level of entanglement via the Schmidt decomposition \cite{Walborn2012}. The Schmidt number, $K = \frac{1}{4}\left(\frac{\sigma_+}{\sigma_-} + \frac{\sigma_-}{\sigma_+}\right)^2$ \cite{law2004}, can be measured from the joint probability distribution $(\text{JPD} = |\psi(x_1, x_2)|^2)$ of the photons using a spatially pixelated photon counter \cite{Hugo2018}. A variant of this DG approximation which considers a Gaussian pump beam with a radius of curvature at the crystal's entry face is presented in \cite{Chan2007}.

\par The most significant property of this wavefunction which is relevant to our work is the exact analytical form of the wavefunction as it propagates$(z)$ \cite{howell2016, Chan2007,Tasca2009}. We define the propagation operator $\left(\hat{R}\left[z,\lambda\right]_{x}\right)$ as a Fresnel propagation integral in one dimension for a monochromatic field of wavelength $\lambda$ \cite{Goodman1996} and is given as:
\begin{equation*}
    \hat{R}\left[z,\lambda \right]_{x}f \equiv \frac{1}{\sqrt{i\lambda z}}\int_{-\infty}^{\infty}e^{\frac{ik(x-x')^2}{2z}}f(x')dx'
\end{equation*}
Then the wavefunction  at $z$ from the crystal center would be given as:
\begin{eqnarray}
\label{eqn:DG prop}
    &\Psi(x_1,x_2;z) =\hat{R}\left[z,\lambda\right]_{x_1} \hat{R}\left[z,\lambda\right]_{x_2} \psi \nonumber \\[6pt] 
 &=\mathcal{N}(z,k)\exp{\left(-\frac{(x_1-x_2)^2}{4(\frac{iz}{k}+\sigma_-^2)}-\frac{(x_1+x_2)^2}{4(\frac{iz}{k}+\sigma_+^2)}\right)}
\end{eqnarray}
where $k = \frac{2\pi}{\lambda}$, with $\lambda$ being the central wavelength of the down-converted photons, and $\mathcal{N}(z,k)$ is a distance-dependent normalization constant. By computing $|\Psi(x_1,x_2;z)|^2$, we see that the probability density remains a DG, with correlation widths given by $\sigma_{\pm}(z) = \sqrt{\sigma_{\pm}^2 + \left(\frac{z}{k\sigma_{\pm}}\right)^2}$\cite{howell2016}. These correlation widths evolve from high correlation in near field(position) to high anti-correlation in far field (momentum) as it propagates whenever $\sigma_+ \gg \sigma_-$  holds. Interestingly, there exists a critical distance,
\begin{equation}
\label{eqn:ploc}
    z_p = \pm k\sigma_+\sigma_-,
\end{equation}
 where $\sigma_+(z_p) = \sigma_-(z_p)$. Here, there exists no correlation in either the position or momentum of the two photons\cite{Tasca2009,Chan2007}. However, this lack of correlation does not imply a loss of entanglement, as the entanglement is preserved throughout the propagation\cite{Chan2007}. This can be verified by performing a Schmidt decomposition at each $z$\cite{Just2013}. Experimentally the preservation of  Schmidt number is shown in Ref.\cite{Reichert2017,Vikas2024}. A closer examination of the wavefunction at $z = z_p$, given as:
\begin{widetext}

\begin{equation}
\label{eqn:DG phase}
    \Psi(x_1,x_2; z_p) = \mathcal{N}(z_p,k)\exp{\left(-\frac{(x_1-x_2)^2}{4\sigma_-(i\sigma_++\sigma_-)}-\frac{(x_1+x_2)^2}{4\sigma_+(i\sigma_-+\sigma_+)}\right)} 
\end{equation}
\begin{equation*}
    = \mathcal{N}(z_p,k) \exp{\left(-\frac{1}{4}\left(\frac{x_1^2+x_2^2}{\sigma_+^2+\sigma_-^2}\right)\right)}\exp{\left(\frac{i}{4\left(\sigma_+^2+\sigma_-^2\right)}\left((x_1-x_2)^2\frac{\sigma_+}{\sigma_-} + (x_1+x_2)^2\frac{\sigma_-}{\sigma_+}\right)\right)},
\end{equation*}
\end{widetext}
reveals that all correlations are transferred to the phase of the wavefunction. Hence even if the complex amplitude can be factorized, the overall quantum state is not separable. Such a state is called a \textit{phase-entangled} biphoton state. It cannot be directly tomographed using spatial photon number correlations. Nevertheless, intensity correlations are useful for detecting entanglement in propagating transverse wavefunctions \cite{Walborn2008}. In our case, we use the vanishing spatial correlations as an indirect signature of the phase-entangled state.
\begin{figure}[!htb]
    \centering
    \includegraphics[width=1.00\linewidth]{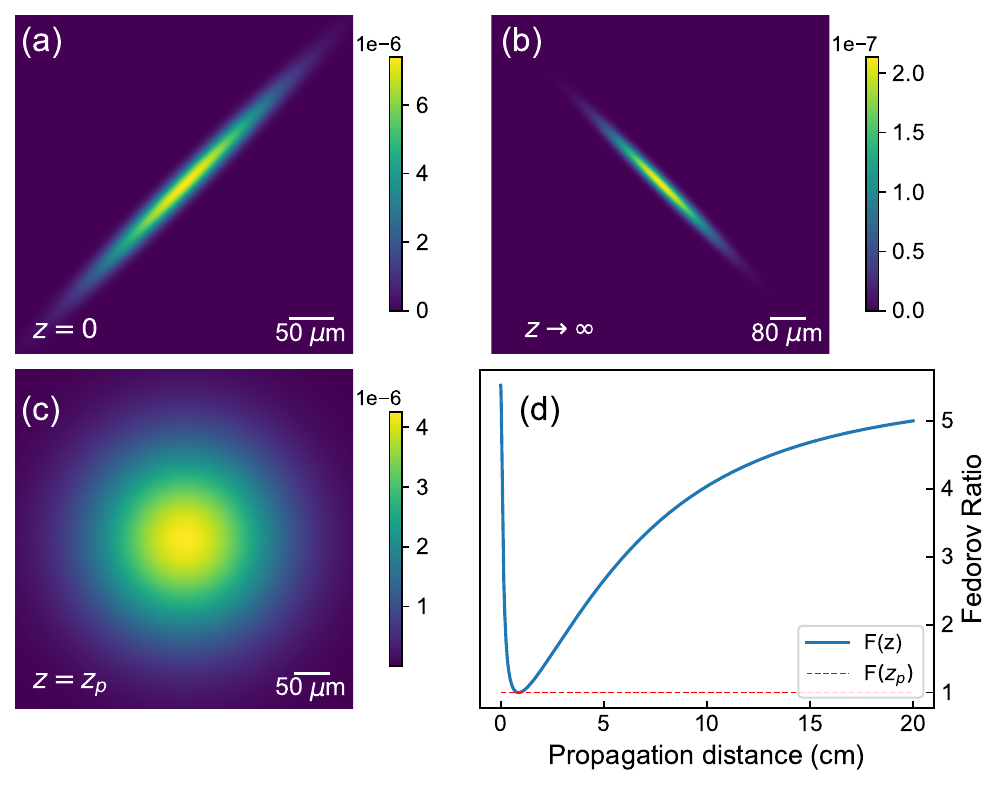}
    \caption{Spatial joint probability density (JPD) of two photons at:  (a) $z = 0$ (b) $z\rightarrow\infty$ (c) $z = z_p$ for a given pair of correlation widths$(\sigma_{\pm})$. (d) Plot of Fedorov ratio as function of $z$. The parameters used for theoretical plots are $\sigma_+ = 140.24~\mu\text{m},\sigma_-=12.56~\mu\text{m}$, taken from experimental estimates and $\lambda = 810\text{ nm}$.}
    \label{fig:FedorovTheory}
\end{figure}

\par Spatial photon number correlations at each $z$ can be characterized by computing the \textit{Fedorov} ratio $(F(z))$. This is done by estimating the joint probability density (JPD) function $\rho(x_1, x_2; z) = |\Psi(x_1, x_2; z)|^2$ and calculating $F(z) = \frac{\sigma_{x_1}}{\sigma_{x_1|x_2}}$ \cite{Fedorov2009}, where $\sigma_{x_1}$ and $\sigma_{x_1|x_2}$ represent the standard deviations of the marginal distribution $\rho(x_1; z)$ and the conditional distribution $\rho(x_1|x_2; z)$, respectively. For the DG approximation, this can be computed as:
\begin{equation}
\label{eqn:FedR}
F(z) = \frac{1}{2}\left(\frac{\sigma_+(z)}{\sigma_-(z)} + \frac{\sigma_-(z)}{\sigma_+(z)}\right) 
\end{equation}
This is illustrated in Fig.\ref{fig:FedorovTheory}, where it is evident that $F(z_p) = 1$. Thus, the Fedorov ratio can serve as an indicator of phase entanglement for the DG biphoton state. In the following section, we describe a simplified experimental setup to study the Fedorov ratio at various $z$. We measure it over the entire propagation domain by folding the distance using a single lens.
\section{Experimental Setup and Analysis}
\label{sec:Exp}
\subsection{Initial state preparation}
To justify our initial theoretical assumption, we need to construct a pump beam with a Gaussian field profile and ensure that the waist of this beam is positioned at the center of the crystal. This process also provides a means to quantify the amount of entanglement, characterized by the Schmidt number, which primarily depends on the size of the pump beam, since the spatial correlation width $\sigma_-$ is determined by fixed crystal parameters. The variation of the quantum state over propagation is achieved through single-lens imaging, which will be detailed in sec. \ref{sec:folding}.

\subsubsection{Gaussian pump beam preparation}
\label{sec: state callibration}
\begin{figure*}[!htb]
    \centering
    \includegraphics[width=0.8\linewidth,trim =1cm 1cm 0.63cm 1cm,clip]{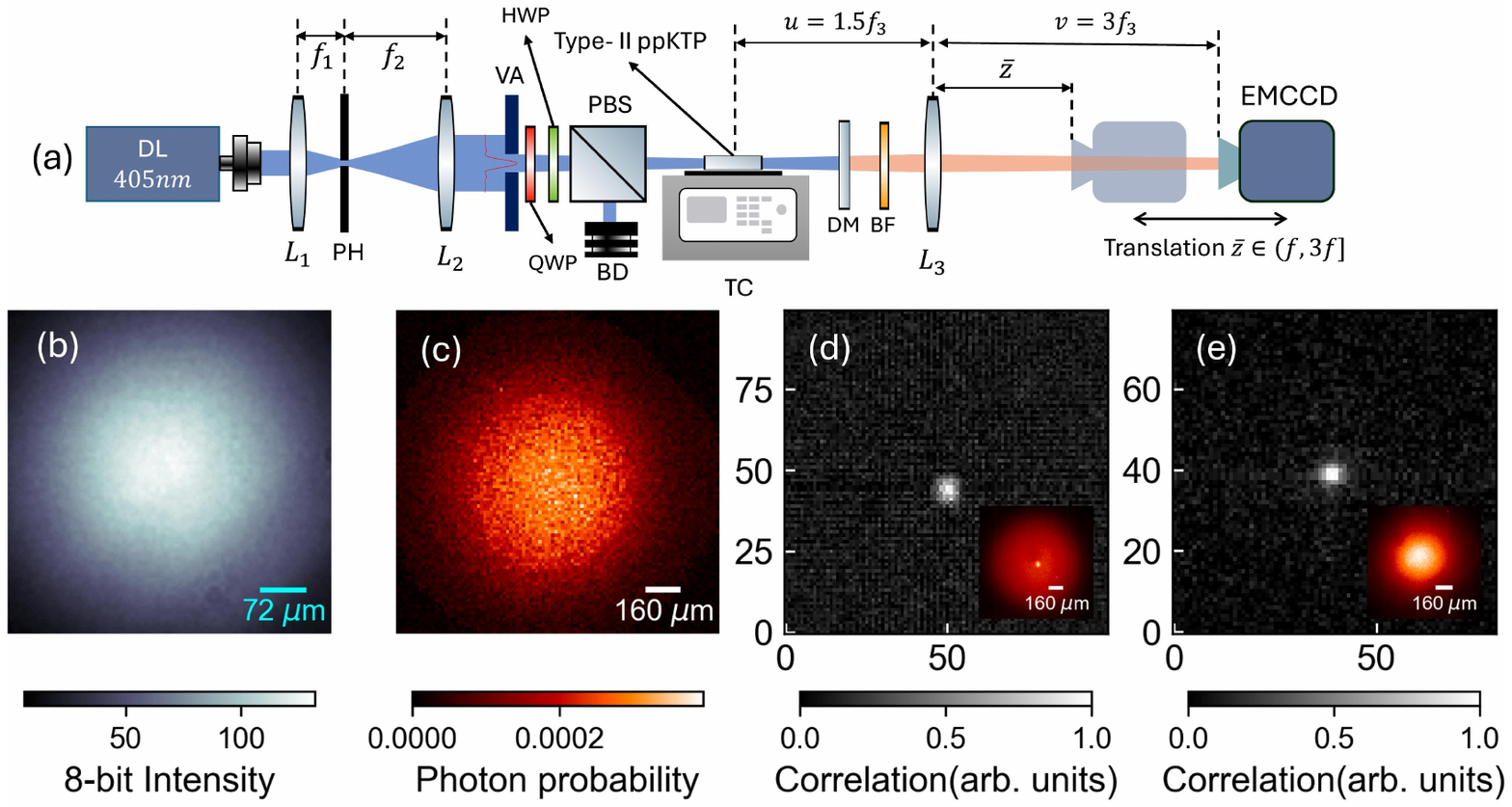}
    \caption{Setup for measuring Fedorov ratio over full range of propagation of the biphoton wavefunction. (a) Schematic of the setup:  DL: $405$ nm Diode Laser  $L_1(f_1 = 35\text{ mm})$, $L_2(f_2 = 75\text{ mm})$, $L_3(f_3 = 40\text{ mm})$ are lenses, PH $=50~\mu$m pinhole, VA: variable aperture. QWP: Quarter-wave plate, HWP: Half-wave plate, PBS: Polarising beamsplitter. TC: temperature controller casing, DM: Dichroic mirror, BF: Band-pass filter. (b) Typical Pump beam obtained on an CMOS camera. (c) One photon marginal distribution for SPDC beam at $\bar{z} = 60\text{ mm}$. (d) Momentum anti-correlation peak at focal plane$(\bar{z}=f_3)$, the scale is max normalized while the inset shows the SPDC beam at this location. (e) Position correlation at the imaging plane $(\bar{z} =3f_3)$, scale is max normalized and inset shows the beam. }
    \label{fig:Setup}
\end{figure*}

 A Gaussian beam is achieved by spatially filtering our pump laser $(\text{Wavelength = }405 \text{nm})$ by focusing it onto a $50\mu m$ pinhole, as shown in the experimental schematic (Fig. \ref{fig:Setup}(a)).  The output beam profile can be well approximated by a Gaussian profile \cite{Zhang2007SpatialFilter}. The location and size of the beam waist are estimated by measuring the beam's propagation as described in Ref. \cite{Vikas2024}. A typical intensity profile is shown in Fig. \ref{fig:Setup}(b). The beam waist location was imaged onto the crystal center using a $4f$ system (not shown in the schematic). The beam waist after the $4f$ setup was measured to be $w_0 \approx 264 \pm 2 ~\mu\text{m}$. We used a type-II periodically poled Potassium Titanyl Phosphate (PPKTP) crystal, with dimensions $10 \text{ mm} \times 2 \text{ mm} \times 1 \text{ mm}$. It was pumped by a horizontally polarized beam and maintained at $40^\circ \text{C}$ to produce near-degenerate, spatially entangled photon pairs at $810~ \text{nm}$. We used a long-pass dichroic mirror (DM) at cut-off  $\lambda=442~ \text{nm}$ and a sharp bandwidth band-pass filter (BF) at $810\pm1.5 ~\text{nm}$ to block the pump photons and filter the entangled photons. We confirmed that the DG approximation was valid by using the sharp BF, in absence of which one expects deviations from the DG approximation.\cite{Vikas2024}.

 \subsubsection{Distance folding using a single lens}
 \label{sec:folding}
To obtain the variation of $F(z)$ over the full range of $z \in \left[0, \infty \right)$, we can \enquote{fold} the domain into a finite region using a single lens, as introduced in \cite{Just2013}. To derive the mapping, we utilize the single-variable operator formalism for coherent optical systems as described in \cite{Goodman1996}. For a field $g(x)$ positioned at a distance $u$ behind a thin lens $L$ of focal length $f$, the output field $\bar{g}(x;\bar{z})$ at a distance $\bar{z}$ in front of the lens is given by:
\begin{equation}
\label{eqn:lensPropagation}
    \bar{g}(x;\bar{z}) = \hat{R}\left[\bar{z},\lambda\right]_{x} \hat{Q}\left[-\frac{1}{f},\lambda \right]_{x}\hat{R}\left[u,\lambda\right]_{x} g 
    \end{equation}
    where $\hat{Q}\left[c,\lambda\right]_{x} g \equiv exp{\left( \frac{i \pi}{\lambda}c x^2\right)} g(x)$ is called a multiplicative quadratic phase, where $c$ has units of inverse length. When a field passes through a thin lens, it obtains a quadratic phase with $c = -\frac{1}{f}$ for a converging lens. We club the three operators in Eq.\eqref{eqn:lensPropagation}  into a single operation  $\hat{\mathcal{L}}\left[u,f,\bar{z}\right]_x$ and simplify it using the commutation relationship between $\hat{Q}$ and $\hat{R}$ .\cite{Goodman1996}
    \begin{eqnarray}
        &\bar{g}(x;\bar{z}) =\hat{\mathcal{L}}\left[u,f,\bar{z}\right]_{x} g   \nonumber \\ [5pt]& =\hat{Q}\left[\frac{1}{\bar{z}-f},\lambda\right]_x\hat{\mathcal{V}}\left[\frac{1}{1-\frac{\bar{z}}{f}}\right]_x\hat{R}[z,\lambda]_x g
    \end{eqnarray}
where  $\hat{\mathcal{V}}\left[s\right]_x g \equiv \sqrt{s}g(s x)$ is a scaling operator with scale length $s$ and $z = u-(\bar{z}f)/(\bar{z}-f)$. Here, the defined $z$ is the propagation distance (in the negative direction) whose wavefunction is mapped at $\bar{z}$. Consequently, one can show that $z\rightarrow -\infty \text{ as } \bar{z}\rightarrow f$ and at this limit the operator $\hat{\mathcal{L}}$  transforms to a simple scaled Fourier transform with a multiplicative phase effectively giving far-field or Fraunhofer diffraction pattern of the input field. While at the imaging distance $\bar{z} = v = (uf)/(u-f)$ from lens, we get  $z = 0$, effectively giving a scaled image with an excess multiplied quadratic phase. Thus the operator $\hat{\mathcal{L}}$ operated for $\bar{z}\in \left(f,v\right]$ maps the propagation of field $g(x;z)$ for $z\in \left(-\infty,0\right]$, with a relative scaling and excess phase. This way we map the entire propagation of $\Psi(x_1,x_2;z)$(Eq.\eqref{eqn:DG prop}). Mathematically we obtain:
\begin{eqnarray}
    &\hat{\mathcal{L}}\left[u,f,\bar{z}\right]_{x_1}\hat{\mathcal{L}}\left[u,f,\bar{z}\right]_{x_2} \psi = \nonumber \\[5pt] & \hat{Q}\left[c,\lambda\right]_{x_1}\hat{\mathcal{V}}\left[s\right]_{x_1} \hat{Q}\left[c,\lambda\right]_{x_2}\hat{\mathcal{V}}\left[s\right]_{x_2} \{ \Psi(x_1,x_2;z)\},
\end{eqnarray}
where $c = 1/(\bar{z}-f) \text{ and } s = 1/(1-\bar{z}/f)$. 

 \par This mapping existing in the domain $z \in \left(-\infty, 0\right]$ is not an issue since the wavefunction $\Psi(x_1, x_2; z)$ maps to its conjugate exhibiting the same physical properties for reversed distances. In fact, the two phase entanglement plane locations are symmetric about the origin as $z_p = \pm k\sigma_+\sigma_-$. Two important points need to be addressed: firstly, the excess phase will not affect the computation of the Fedorov ratio, as all calculations are based on the spatial probability density of the two photons involving only the amplitude part. Secondly, the scaling factor will not influence the result since the Fedorov ratio is a \textit{ratio} and remains invariant. A different theoretical approach using Fractional Fourier Transform can be found in Ref~\cite{Tasca2008}.

\par In our experiments, we choose $f = f_3 = 40\text{ mm}$ and place the lens at a distance of $1.5f = 60\text{ mm}$ from the crystal center, as shown in Fig. \ref{fig:Setup}(a). This effectively folds the domain to $\bar{z} \in \left(f, 3f \right] = (40\text{ mm}, 120\text{ mm}]$, between the focal plane and the imaging plane. To experimentally establish these planes, we follow the method described in \cite{Edgar2012}, which demonstrates momentum anti-correlation in the focal plane and position correlation in the imaging plane. This is achieved by capturing multiple frames of photon pairs on an Electron Multiplying Charge Coupled Device (EMCCD), using it as a spatial photon counter. 
\par We acquire 10,000 (20,000) frames at the focal plane (imaging plane), and employ our recently-developed method for photon counting that involves multi-thresholding of the EMCCD counts\cite{Rounak2024}. The acquired frames are auto-convolved (auto-correlated) and averaged, to measure the anti-correlation (correlation) at the focal (imaging) plane. Background correction is applied to remove classical noise. Fig. \ref{fig:Setup}(d) shows the anti-correlation peak, which is fitted with a Gaussian to yield $\sigma_+ \approx 2.34$ pixels. Given the EMCCD pixel size of $16~\mu$m, the focal length $f_3 = 40\text{ mm}$, and the central wavelength of SPDC photons is $810\text{ nm}$, we estimate $\sigma_+ \approx 140.2~\mu$m. Fig. \ref{fig:Setup}(e) shows the correlation peak leading to $\sigma_- \approx 12.6~\mu$m. Hence, the phase entanglement plane location is  calculated as $z_p  \approx 13.7\text{ mm}$ from the crystal center using Eq.\eqref{eqn:ploc}, with the Schmidt number $K \approx 31$. The phase entanglement plane in the folded domain is at $\bar{z}_p\approx91.1\text{ mm}$ from the lens.  

\subsection{Measurement and filtering of biphoton JPD }
\subsubsection{Experimental measurement of JPD}
\label{sec:computing JPD}
To estimate the Fedorov ratio at each $\bar{z}$, we construct the JPD.  We model the complete 4-variable $(x_1,y_1;x_2,y_2)$ JPD for any biphoton wavefunction incident on a pixelated photon-counting camera. If each pixel counts at the most one photon, $c_{i} = \{0,1\}$, where $i$ represents pixel centered at $(x_{i},y_{i})$, then the discrete 4-dimensional JPD $\Gamma_{ij}$ is given exactly as \cite{Hugo2018}:
\begin{equation}
\label{eqn:hugo}
    \Gamma_{ij} = \frac{1}{2\eta^2\mu}\text{ln}\left(1+\frac{\langle c_ic_j\rangle - \langle c_i\rangle\langle c_j\rangle}{(1-\langle c_i\rangle)(1-\langle c_j\rangle)}\right),
\end{equation}
where $\eta$ is the quantum efficiency of the camera, $\mu$ is the mean photon-pair rate (assumed to be Poisson distributed in time) and $\langle \cdot \rangle$ indicates ensemble average over the frames. We ensure that the mean photon number per pixel per exposure  is  $\leq0.1$ for Eq.\eqref{eqn:hugo} to hold.  We use our method \cite{Rounak2024} to construct photon threshold map over the pixels to perform single photon counting. The ensemble averages over $M$ frames is then given by \cite{Hugo2019}:
\begin{eqnarray}
\label{eqn:stats}
    &\langle c_ic_j \rangle = \frac{1}{M}\sum_{k=1}^M {c_i^{(k)}c_j^{(k)}} \nonumber \\[6pt]&\langle c_i \rangle \langle c_j \rangle \approx \frac{1}{(M-1)} \sum_{k=1}^{M-1} {c_i^{(k)}c_j^{(k+1)}} 
\end{eqnarray}
where $c^{(k)}_{i}$ is the $i^\text{th}$  pixel in the $k^{\text{th}}$ frame. The second equation in  Eq.\eqref{eqn:stats} is the product of independent frame averages. This is an approximate expression because the full solution consists of  all possible pairings of frames \cite{Hugo2019}. Nonetheless, it is acknowledged to be a good approximation to first order. Although  Eq.\eqref{eqn:stats} leads to some statistical noise discussed in the next sec. \ref{sec:bloom and background}, it's computationally optimal to use on large datasets we acquire. Experimentally, we compute $\Gamma_{ij}$ using equation Eq.\eqref{eqn:hugo} which is then reduced to a two-dimensional distribution $\rho(x_i,x_j;\bar{z})$ with one coordinate of each pixel as
\begin{equation}
\label{eqn:reduction}
\rho(x_i,x_j;\bar{z}) = \sum_{y_i,y_j}{\Gamma_{x_i,y_i;x_j,y_j}}    
\end{equation}
\begin{figure}[!htbp]
    \centering
    \includegraphics[width=1.0\linewidth,trim =1.5cm 1.5cm 1.5cm 1.5cm,clip]{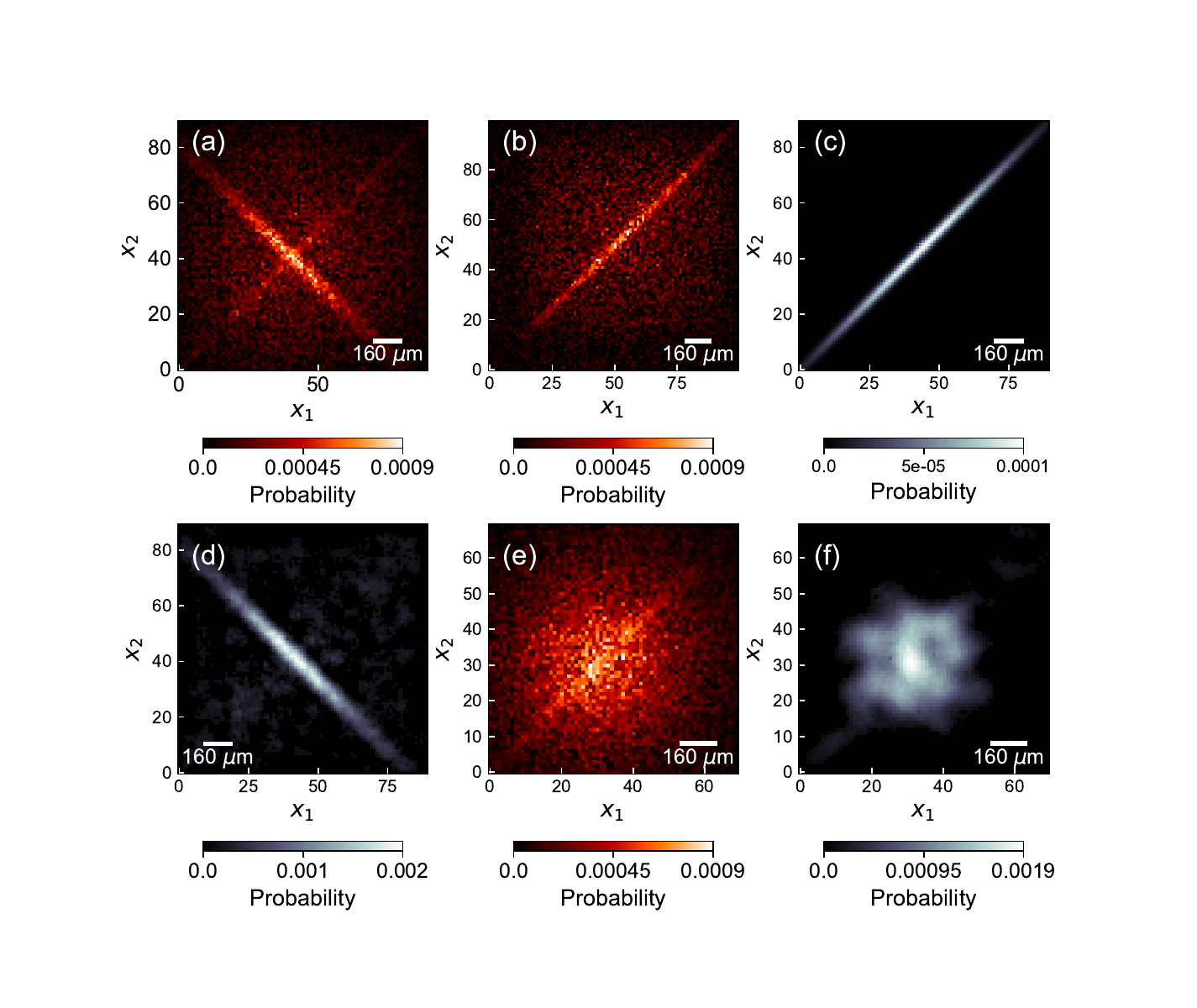}
    \caption{Depiction of computed two variable JPDs $\rho(x_1,x_2;\bar{z})$ and filtering of data. All the axes are in pixel units (pixel size $=16~\mu\text{m}$) (a) Unfiltered $\rho(x_1,x_2;\bar{z} = f_3)$ obtained using Eq.\eqref{eqn:hugo} and Eq.\eqref{eqn:reduction}. (b) JPD obtained from uncorrelated source to find correlation among blooming pixels. (c) The  $\rho_{\text{bloom}}$ constructed for baseline correction of Fig. \ref{fig:filter}(a) (see appendix \ref{app:bloom}) (d) Post bloom-correction and Band-pass Fourier filtering of Fig.  \ref{fig:filter}(a). (e) Unfiltered $\rho(\bar{z} = 100~\text{mm})$ (f) Filtered Fig. \ref{fig:filter} (e).}
    \label{fig:filter}
\end{figure}

Using a single transverse spatial coordinate is justified since a DG has azimuthal symmetry. This step also helps us visualize the JPD, as evident from Fig. \ref{fig:filter}. A raw computation of $\rho(x_i,x_j;\bar{z}=f_3)$ for $2\times10^5$ frames is shown in Fig. \ref{fig:filter}(a). We primarily observe a bright anti-diagonal line representative of momentum anti-correlation, which constitutes the signal. The rest of the image in the frame constitutes the noise, of which two sources are identified. Firstly, here, a diagonal line arises from the blooming on the EMCCD camera. Secondly, the unavoidable room light scatter also leads to background noise in the correlation image. The elimination of both is discussed in the next section.  

\subsubsection{Correcting for camera and statistical noise}
\label{sec:bloom and background}
Blooming is a phenomenon that occurs in all types of charge-coupled devices (CCDs), where a pixel detecting a photon sometimes deposits charge on an adjacent pixel. This can happen due to the finite charge capacity of individual pixels or when a pixel reaches its maximum charge transfer rate \cite{Blooming_Nasir2017}. In our case, the electron well-depth of a pixel is deep enough to avoid the first situation, but blooming does occur when charges are transferred rapidly across pixels during single-photon counting. While this can be mitigated by lowering transfer rates, it significantly impacts acquisition time. Blooming in the camera manifests as a false correlation among pixels and is prominent in a pixel looking at higher mean photon rate than others. Computing a pixel-wise correlation for an uncorrelated source shows that blooming exactly behaves like a DG (Fig~\ref{fig:filter}(b)), as elaborated in Appendix \ref{app:bloom}. For this reason, blooming is known to contaminate photon pair correlation in the imaging plane. To correct for this, we theoretically estimate a DG representing blooming $\rho_{\text{bloom}}$ at each $\bar{z}$ that serves as a baseline correction for $\rho(x_1,x_2;\bar{z})$. For example, the correction $\rho_{\text{bloom}}$ for  Fig. \ref{fig:filter}(a) is given in Fig. \ref{fig:filter}(c). 

 A Butterworth band-pass filter was used to counter statistical noise due to finite sampling. The Bloom-corrected and filtered data obtained from Fig~\ref{fig:filter}(a) is shown in Fig~\ref{fig:filter}(d). Identical filtering parameters and methodology was utilized for filtering data at each $\bar{z}$. A typical filtering example for $\bar{z} = 100\text{ mm}$, corresponding to the unfiltered JPD of Fig~\ref{fig:filter}(e) is shown in Fig~\ref{fig:filter}(f).
\section{Results}
\label{sec:results}
\begin{figure}[!htbp]
    \centering
    \includegraphics[width=1.\linewidth,trim =1.5cm 1.5cm 1.5cm 1.5cm,clip]{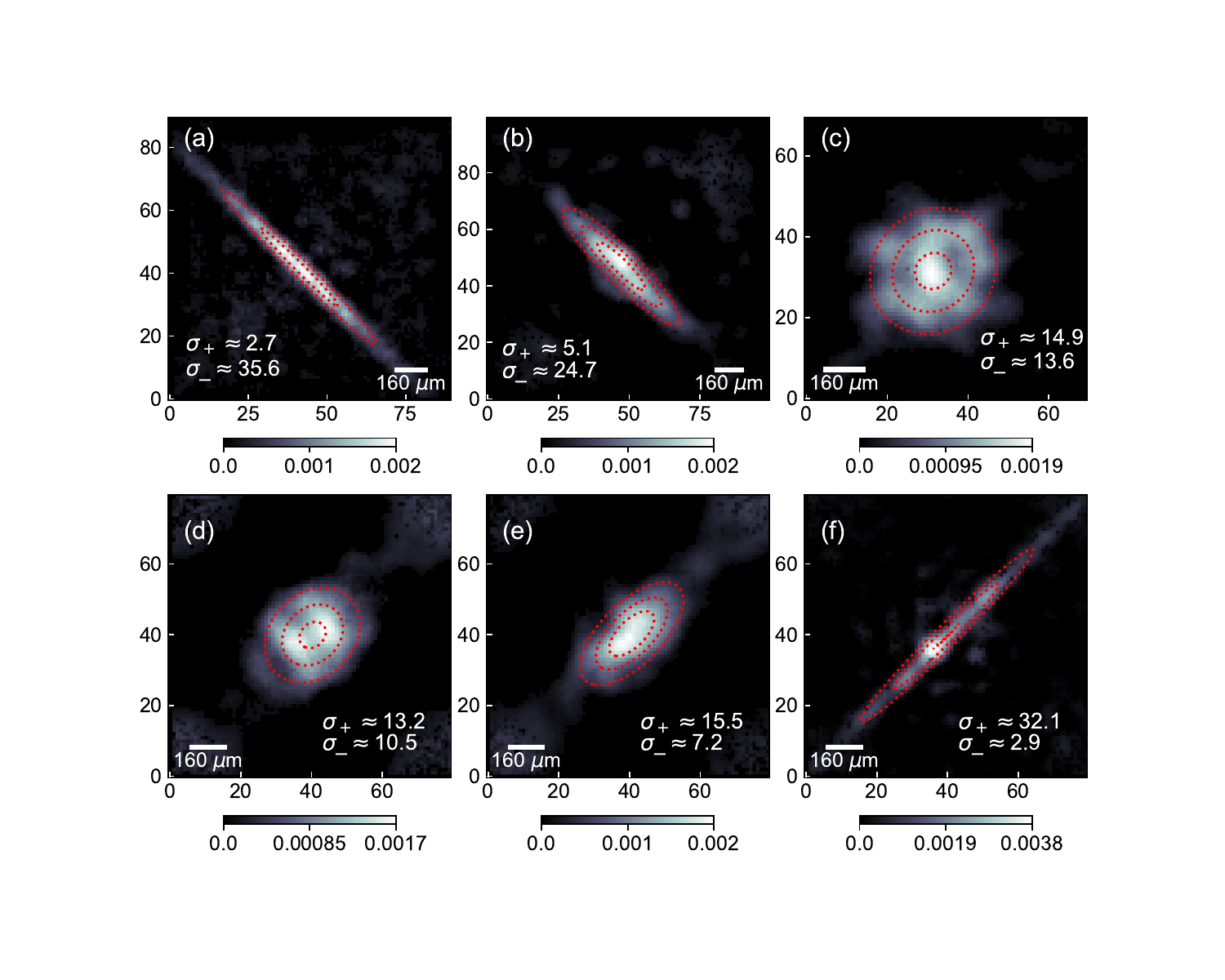}
    \caption{Filtered two variable JPDs $\rho(x_1,x_2;\bar{z})$ at $\bar{z} = $ (a) $f_3 $ (b) $65$ mm (c) $85$ mm (d) $100$ mm (e) $115$ mm (f) $3f_3$. All the axes are in pixel units (pixel size $=16~\mu\text{m}$)  }
    \label{fig:jpd}
\end{figure}

Finally, the noise-corrected images are shown in Fig. \ref{fig:jpd}. Six images corresponding to different $\bar{z}$ are illustrated. The red dotted contours represent the fit of the DG (Eq.\eqref{eqn:DG}), from which $\sigma_\pm(\bar{z})$ are estimated. The fit parameters are mentioned in the legend. The swapping of $\sigma_+$ and $\sigma_-$ beyond Fig~\ref{fig:jpd}(c) shows that the photon-pair correlation changes from momentum anti-correlation to position correlation.  For $\bar{z}= 85~$mm, the $\sigma_-$ and $\sigma_+$ are within $10\%$ of each other. The plane at Fig~\ref{fig:jpd}(c) thus marks the plane closest to maximal phase entanglement. A more detailed profile analysis of these images is shown in Appendix~\ref{app:DG}.

\begin{figure}[!htbp]
    \centering
    \includegraphics[width=0.85\linewidth,trim = 1cm 0cm 1.5cm 1cm,clip]{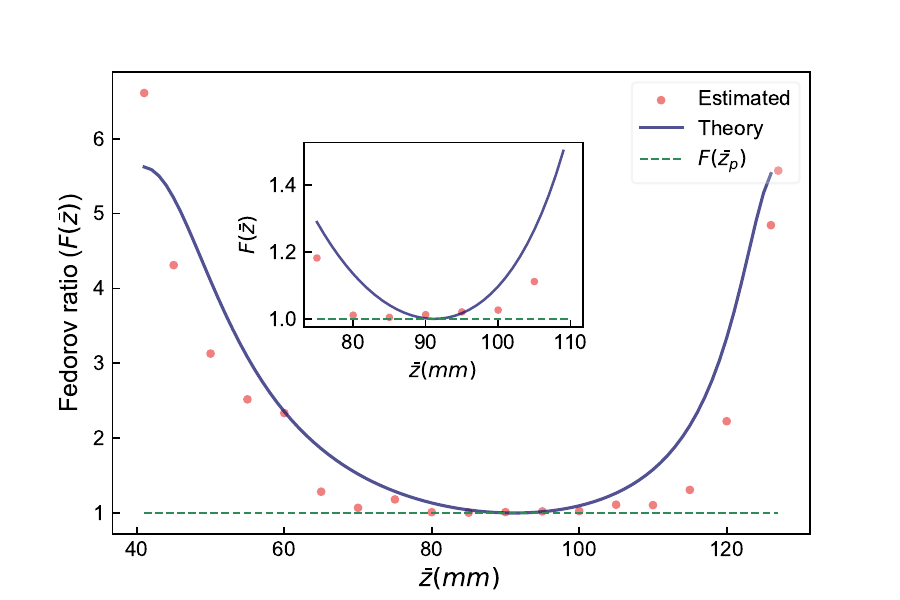}
    \caption{Experimental variation of the Fedorov ratio (orange dots) with folded distance $(\bar{z})$, along with theoretically estimated curve (blue). The green dashed line indicates a Fedorov ratio of 1.}
    \label{fig:Fz_vs_z}
\end{figure}

To assess the agreement of our measurements with theoretical predictions, we plot the computed Fedorov ratio at specific folded distances, as indicated by the orange points in Fig.~\ref{fig:Fz_vs_z}. As mentioned in sec.\ref{sec:folding}, the values $(u,f,\sigma_-,\sigma_+) = (60~\text{mm, }40\text{ mm, }12.6~\mu \text{m}, 140.2~\mu \text{m})$ were used. The theoretical variation with folded distance $\bar{z}$ is evaluated using $z = u-(\bar{z}f)/(\bar{z}-f)\text{, and }\sigma_{\pm}(z) = \sqrt{\sigma_{\pm}^2 + \left(\frac{z}{k\sigma_{\pm}}\right)^2}$ and Eq.\eqref{eqn:FedR}. The calculated curve is shown in blue. The experiment accurately captures the expected theoretical trend, with the experimental Fedorov ratio approaching unity near the point of phase entanglement at $\bar{z} \approx 91~\text{mm}$. Evidently, the measured data at $F(\bar{z})$ close to 1 are rather flat. This is primarily due to two reasons. Firstly the low-frequency background noise arising during the computation of the JPD overestimate the values of $\sigma_{\pm}(\bar{z})$. Second the propagation of the bi-photon wavefunction doesn't exactly follow DG propagation in practical scenarios, and the considerations here are approximates\cite{Reichert2017}. A true simulation of this propagation can be obtained by considering the true wavefunction (Eq. \ref{eqn:SPDC original}). Nonetheless, the loss of spatial correlation signifies the wavefunction is correlated in phase demanded by conservation of entanglement. This corresponds to the situation where accidental spatial coincidences are comparable to the true coincidences between photon pairs, and underlines the challenges in measuring the plane of maximal phase entanglement. The inset of Fig.~\ref{fig:Fz_vs_z} emphasizes the minimum point.

\section{Interference experiments}
Interference is by far the simplest yet indispensable tool for measuring phase structures in waves, whether in classical or quantum mechanical scenarios, since it is known to ``mix" phase fronts and translate them into intensity variations in the far field. The key focus here is on \textit{phase correlations among photon pairs}; therefore, beyond merely observing interference effects, we require a mathematical framework that can isolate intensity correlations arising purely from entanglement in the photon pairs, discounting marginal contributions from symmetry effects.
\par We employ the concept of symmetric excess fourth-order correlation \cite{Saleh2000}, which functionally measures \textit{pure} two-photon intensity correlations at any plane of interest (details in sec.\ref{subsec:DS Theory}). Our core strategy is to illuminate a double slit (DS) with two-photon phase-entangled light (which we use interchangeably with phase-correlated light) and observe the results in far field. We will theoretically and experimentally demonstrate that the far-field two-photon probability distribution changes significantly due to the presence of a correlated phase front. Moreover, it is known that amplitude correlations in the illuminating quantum state are responsible for pure two-photon intensity correlations due to far-field interference, as documented in reported experiments \cite{Monken1999,Brida2003}. However, since a phase-entangled state is initially amplitude-uncorrelated, the only plausible explanation for observing far-field intensity correlations for phase entangled light lies in the inherent phase correlations. To serve both as a control and to highlight the contrasting physical scenarios of far-field intensity distributions arising from amplitude versus phase correlations, we also perform the same two-photon interference experiment using position-correlated light, which is essentially amplitude-correlated.

\subsection{Theoretical analysis and experimental setup}
\label{sec:DS setup and theory}
\subsubsection{Theory}
\label{subsec:DS Theory}
\par Mathematically, the DS aperture function can be modelled in 1D as\cite{Abouraddy2001_DS}:
\begin{equation}
    \label{eqn: DS_func}
    \text{DS}(x,d,a) = \text{Box}\left(\frac{x-\frac{d}{2}}{a}\right)+\text{Box}\left(\frac{x+\frac{d}{2}}{a}\right)
\end{equation}
where $d$ is slit separation, $a$ is the slit width, while $\text{Box}(x) =\begin{cases} 0,&|x|>\frac{1}{2}\\ 1,&|x| \leq  \frac{1}{2} \end{cases} $, is a unit box function. If $\psi_{DS}$ is the biphoton state illuminating the DS and $f_3$ is the focal length of the Fourier-transforming lens mapping the interference in far-field, then the observed two-photon probability density (from the wavefunction $\psi_{\text{inter}})$ is given as:
\begin{widetext}
\begin{equation}
\label{eqn: DS density}
    \rho_{\text{inter}}\left(x_1,x_2\right) =\left|\psi_\text{inter}(x_1,x_2)\right|^2 = \left| \hat{\mathcal{V}}\left[\frac{1}{\lambda f_3}\right]_{x_1}\hat{\mathcal{V}}\left[\frac{1}{\lambda f_3}\right]_{x_2}\mathcal{F}_{x_1,x_2}\left(\text{DS}(x_1,d,a)~\text{DS}(x_2,d,a)~\psi_{DS}\right)\right|^2  
\end{equation}
\end{widetext}
where $\mathcal{F}_{x_1,x_2}$ is the two-variable Fourier transform and $\hat{\mathcal{V}}$ is the scaling operator (sec.\ref{sec:folding}).  Thus, substituting $\psi_{DS}$ with Eq.~\ref{eqn:DG}  gives interference due to position-correlated light, while substituting with  Eq.~\ref{eqn:DG phase} gives interference due to phase-correlated light. The forms of these densities with experimental schematics are shown in the next section (sec.\ref{subsec:sim and exp}, Fig.\ref{fig:ImgPl interference setup} and Fig.\ref{fig:PhasePl interference setup}, respectively). To mathematically formalize the role of two-photon entanglement (manifesting as amplitude or phase correlations) in either wavefunction densities or interference density $\rho_{\text{inter}}$ (Eq.\ref{eqn: DS density}), we introduce the concept of excess symmetric (or auto) fourth-order correlation function \cite{Saleh2000}, given as:

\begin{equation}  
\label{eqn:delG2}  
    \small \Delta G^{(2)}(x_1,x_2) = G^{(2)}(x_1,x_2) - G^{(1)}(x_1,x_1)G^{(1)}(x_2,x_2)  
\end{equation} 

where $G^{(2)}(x_1,x_2)  = |\psi(x_1,x_2)|^2 = \rho(x_1,x_2)$, $\psi$ and $\rho$ are the wavefunction and it's corresponding density of the state under study. $G^{(1)}(x_i,x_i) = \int{G^{(2)}(x_1,x_2)\text{ d}x_j}, \quad i,j=\{1,2\} \text{ and } i \ne j$. For a  symmetrical system (such as our case), one can show that  $G^{(1)}(x_i,x_i) = \rho_{\text{marginal}}(x_i) = \int \rho(x_1,x_2) dx_1 = \int \rho(x_1,x_2) dx_2$. This function $\Delta G^{(2)}(x_1,x_2)$ explores the role of two-photon coherence by quantifying pure amplitude correlations among biphotons. This is achieved by eliminating \textit{classical} correlations (extracted using marginal distributions) from the probability densities. In other words, since the two-photon density carries information about both entanglement and the one-photon wave, if we subtract the classical correlation function—extracted from the marginals and extended to the two photon space by performing as simple outer product as $G^{(1)}(x_1,x_1)G^{(1)}(x_2,x_2)$—what remains is the contribution arising purely from entanglement between the photons.
\par For example, if we compute $\Delta G^{(2)}$ for the position-correlated light characterized by the DG state (Eq.\ref{eqn:DG}, Fig.\ref{fig:FedorovTheory}(a)), we would observe that it is almost functionally identical to the density. This indicates that all of the entanglement manifests as amplitude correlations in this state. In contrast, if we use the phase-correlated DG wavefunction (Eq.\ref{eqn:DG phase}, Fig.\ref{fig:FedorovTheory}(c)) to compute $\Delta G^{(2)}$, we find that it evaluates to zero, signifying the absence of two-photon amplitude correlations. At this point, $\Delta G^{(2)}$ conveys the same information as the Fedorov ratio (sec.\ref{sec:theory}, Eq.~\ref{eqn:FedR}). However, the inferences become far more interesting and distinct when we compute $\Delta G^{(2)}$ using the interference density $\rho_{\text{inter}}$ computed using the above mentioned states. Thus $\rho_{\text{inter}}$ for position-correlated light is obtained by setting $\psi_{DS}$ in Eq.\ref{eqn: DS density} to the DG wavefunction given by Eq.\ref{eqn:DG}, while for phase-correlated light, it can be obtained by setting $\psi_{DS}$ to phase entangled state given by Eq.\ref{eqn:DG phase}.
\subsubsection{Experimental schematics, simulations and results}
\label{subsec:sim and exp}
\begin{figure}[!htbp]
    \centering
    \includegraphics[width=0.9\linewidth,trim = 0.5cm 0.25cm 0.5cm 0.75cm,clip]{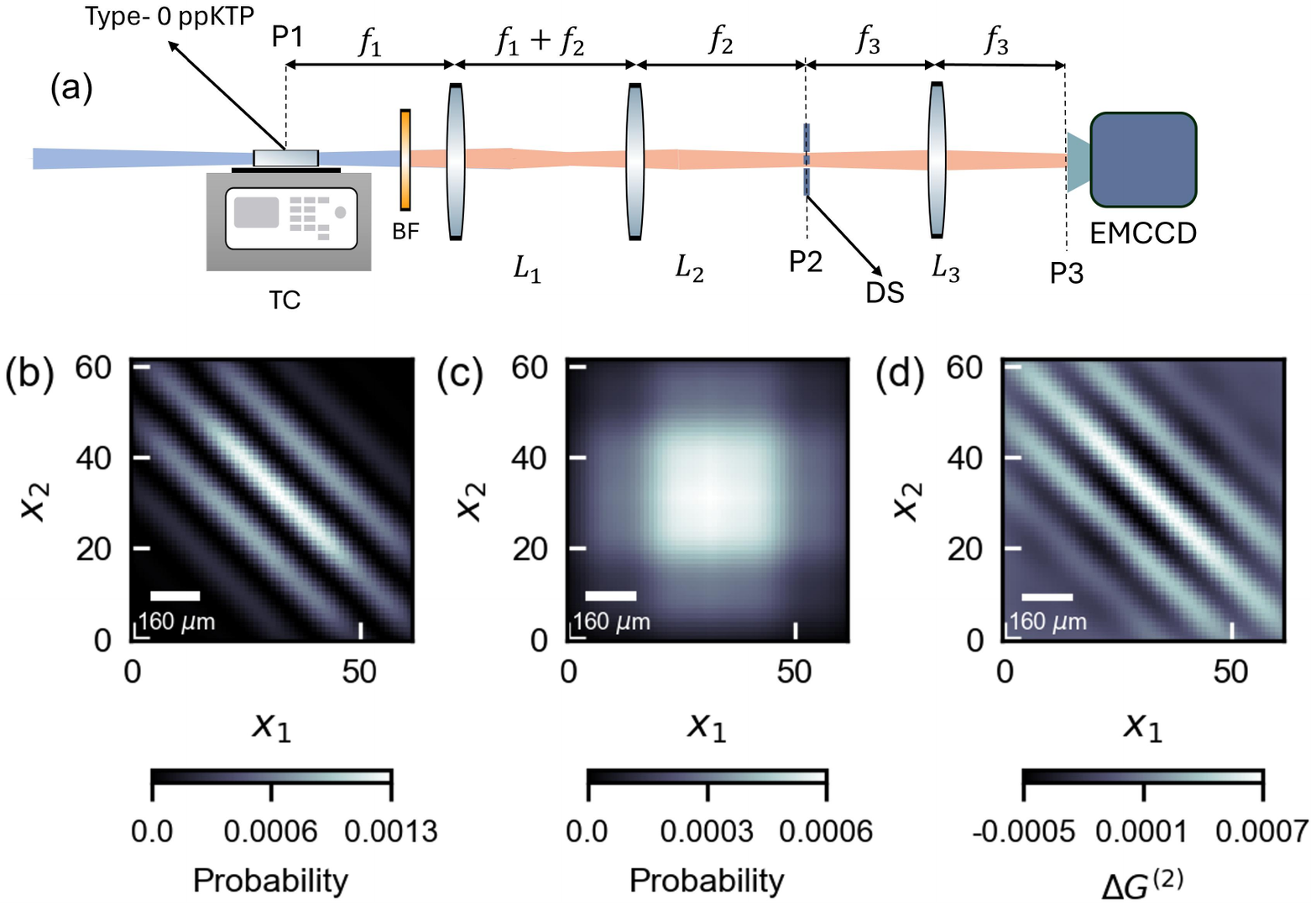}
    \caption{Double slit interference (slit width = $150~\mu$m, slit separation = $400~\mu$m) using position correlated light. (a) Schematic details before crystal are same as Fig.\ref{fig:Setup}. P1 $= $ crystal center, P2 $=~4F$ imaging onto DS (lenses $L_1~(f_1 = 75~\text{mm})\text{, }L_2~(f_2 = 150~\text{mm}))$, P3 $=$ focal plane of $L_3~(f_3 = 125~\text{mm})$. (b) Simulated $\rho_{\text{inter}}(x_1,x_2)$ (Eq.\ref{eqn: DS density}) using $\psi_{DS}=$ Eq.\ref{eqn:DG}. (c) Simulated classical correlation map $G^{(1)}(x_1,x_1)G^{(1)}(x_2,x_2)$ using marginal of $\rho_{\text{inter}}$. (d) Simulated $\Delta G^{(2)}(x_1,x_2)$ (Eq.\ref{eqn:delG2}) using $\rho_{\text{inter}}$. Essentially this is obtained by subtracting (c) from (b).}   
    \label{fig:ImgPl interference setup}
\end{figure}

\begin{figure}[!htbp]
    \centering
    \includegraphics[width=0.9\linewidth,trim = 0.5cm 0.25cm 0.5cm 1cm,clip]{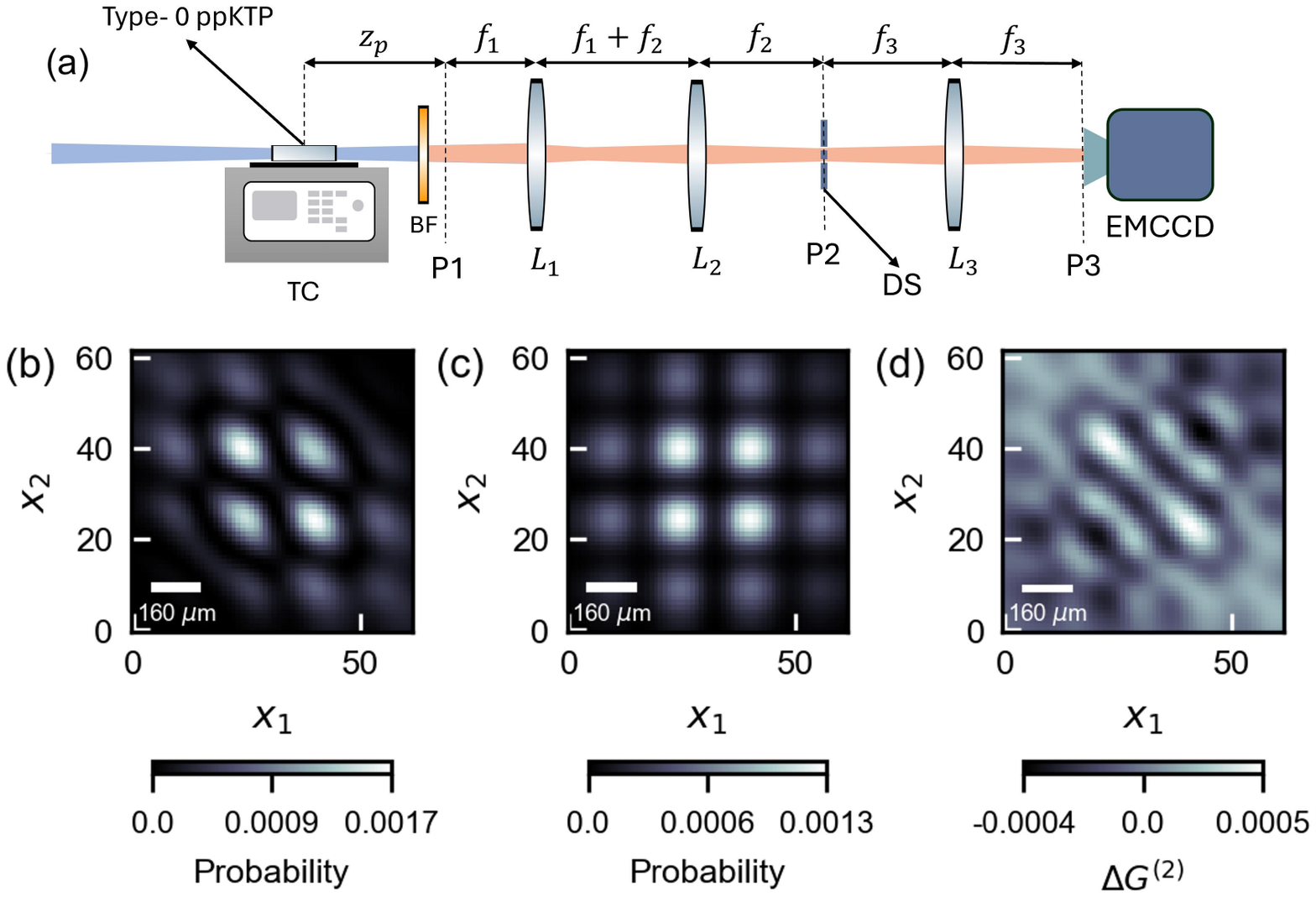}
    \caption{Double slit interference (slit width = $150~\mu$m, slit separation = $400~\mu$m) using phase correlated light. (a) Schematics just before the crystal are same as Fig.\ref{fig:Setup}. P1 $= $ phase entanglement plane ($z_p \approx 2.3$ cm), P2 $= ~4F$ imaging onto DS (lenses $L_1~(f_1 = 75~\text{mm})\text{, }L_2~(f_2 = 150~\text{mm}))$, P3 $=$ focal plane of $L_3~(f_3 = 125~\text{mm})$. (b) Simulated $\rho_{\text{inter}}(x_1,x_2)$ (Eq.\ref{eqn: DS density}) using $\psi_{DS}=~$ Eq.\ref{eqn:DG phase}. (c) Simulated classical correlation map $G^{(1)}(x_1,x_1)G^{(1)}(x_2,x_2)$ using marginal of $\rho_{\text{inter}}$. (d) Simulated $\Delta G^{(2)}(x_1,x_2)$ (Eq.\ref{eqn:delG2}) using $\rho_{\text{inter}}$. Essentially this is obtained by subtracting (c) from (b).}   
    \label{fig:PhasePl interference setup}
\end{figure}

As seen in sec.\ref{sec:folding}, the single-lens mapping leads to excess phase fronts in the wavefunction, which is undesirable while studying interferences. Hence, all planes where the wavefunctions of interest exist, i.e., the plane at the crystal center for position-correlated light (Eq.\ref{eqn:DG}) and the plane at a distance $z_p$ from the crystal center for phase-correlated light, are directly mapped onto the DS using a $4F$ setup. This is followed by obtaining the far-field two-photon field using a single Fourier-transforming lens and statistically deducing densities from multiple frames $(500,000)$ acquired on the EMCCD camera, using the method described in sec.\ref{sec:computing JPD}. The corresponding experimental schematics are shown in Fig.\ref{fig:ImgPl interference setup}(a) and Fig.\ref{fig:PhasePl interference setup}(a), respectively.

\par We employed a type-0 crystal for these experiments to generate same-polarization photons, in order to rule out any role of polarization in the observed phenomena, and for its inherently high brightness, to compensate for the photon loss after passing through a DS. High spatial entanglement is attained using a wider pump beam and a smaller crystal of length $5$ mm. Following the methods described in sec.\ref{sec: state callibration}, we experimentally determine the values of $(\sigma_+,\sigma_-) \approx (326 ~\mu \text{m},~9~\mu\text{m})$ along with the approximate location of the phase entanglement plane at $z_p \approx 2.3~$cm.  To experimentally verify the fidelity of the phase-correlated source, photon counting data was collected after $4F$ imaging of the phase plane at $z_p$ and the Fedorov ratio was computed using the same procedure described in sec.\ref{sec:computing JPD} and sec.\ref{sec:results}. The estimated value turned out to be $\approx$ 1.01, which is just  $1\%$ away from the true value of unity. This shows that the state was well uncorrelated in amplitude and we're accurately looking at phase-correlated light.

\par
\begin{figure}[!htbp]
    \centering
    \includegraphics[width=1.0\linewidth, trim = 1cm 0.1cm 1.2cm 1cm,clip]{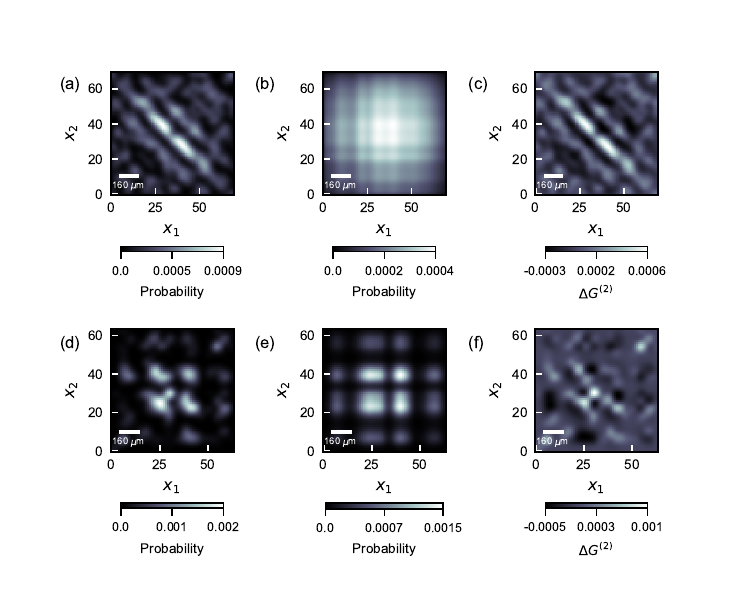}
    \caption{Experimentally obtained \{(a),(d)\} JPD (sec.\ref{sec:computing JPD}) $\rho_{\text{inter}}(x_1,x_2)$ (Eq.\ref{eqn: DS density}), \{(b),(c)\} Marginal images $G^{(1)}(x_1,x_1)G^{(1)}(x_2,x_2)$, \{(c),(f)\} Pure two-photon correlation $\Delta G^{(2)}(x_1,x_2)$ (Eq.\ref{eqn:delG2}) for position correlated light (Eq.\ref{eqn:DG}) and phase correlated light (Eq.\ref{eqn:DG phase}) respectively. }
    \label{fig:DS exp plots}
\end{figure}

 Using $\sigma_\pm$ and procedure described in sec.\ref{subsec:DS Theory}, the simulated forms of densities $\rho_{\text{inter}}$ (considering the magnification of the $4F$ system) for interference using position-correlated light are shown in Fig.\ref{fig:ImgPl interference setup}(b), and for phase-correlated light in Fig.\ref{fig:PhasePl interference setup}(b). The experimentally obtained densities are shown in Fig.\ref{fig:DS exp plots}(a) and Fig.\ref{fig:DS exp plots}(d), respectively. As expected, both of them show interference fringes in the two-photon densities arising from some kind of correlations in the illuminating source.

\par To infer the type of quantum correlations, we first look at position-correlated light. If we consider the marginal, which is essentially treating position-correlated light as a single-photon source, it will behave like an incoherent source \cite{Monken1999,Walborn2010}, since the pairs of photons can appear randomly at any point on the DS. Hence, the simulated 1-photon correlation image, as shown in Fig.\ref{fig:ImgPl interference setup}(c), as well as the experimentally obtained image in Fig.\ref{fig:DS exp plots}(b), shows a flat feature. When we compute $\Delta G^{(2)}(x_1,x_2)$ (Eq.\ref{eqn:delG2}) using this and the density—as shown in simulation by Fig.\ref{fig:ImgPl interference setup}(d) and experimentally in Fig.\ref{fig:DS exp plots}(c)—we see it is functionally similar to the two-photon density, a feature also shown by the DG itself. This shows that coherence is only induced (hence interference is observed) when both photons are simultaneously measured. Thus, one can conclude that the interference is solely mediated by the inherent amplitude correlation of the DG due to entanglement, with no classical contribution.

\par In a similar spirit, when we consider phase-correlated light—the marginal for each photon is a Gaussian beam with no amplitude cross-correlation among them—thus, individually, each photon behaves like an independent Gaussian beam, with classical coherence appearing from the structure of the beam. Essentially, one can consider the analogy of a single-photon Gaussian beam as large as the DS itself incident on it. As intuition would suggest, we should see independent one-photon interference fringes in the marginal correlation image, which we readily observe in simulation, as shown in Fig.\ref{fig:PhasePl interference setup}(c), and also in the experimental data in Fig.\ref{fig:DS exp plots}(e). Thus, one might expect $\Delta G^{(2)}(x_1,x_2)$ computed using the above marginal image and the interference density from phase-correlated light to be identically zero, as is the case with the phase-entangled state. But on the contrary, it shows a \textit{non-zero} structure both in simulation, as shown in Fig.\ref{fig:PhasePl interference setup}(d), and in experiment, as in Fig.\ref{fig:DS exp plots}(f). This tells us, without ambiguity, that there is pure two-photon correlation present in the quantum state after interference. Since the phase-correlated source illuminating the DS does not have any two-photon amplitude correlation to begin with, the only way $\Delta G^{(2)}\ne 0$ after interference is that two-photon phase correlation manifests as pure two-photon amplitude correlation in the far-field due to the interference effect. In a sense the interference effect converts the inherent phase correlation into observable pure two photon amplitude correlation.
\par This experimental phenomenon of excess two-photon correlation after interference from the DS illuminated by phase-entangled light serves as direct proof that entanglement is manifesting as phase correlations in the quantum state. The experimental data for the marginal distribution (Fig.\ref{fig:DS exp plots}(e)) shows a slight bias in its distribution of interference lobes (not seen in simulation), due to initial alignment errors of the DS with respect to the center of the phase-correlated beam. For the same reason, the experimentally obtained $\Delta G^{(2)}$ (Fig.\ref{fig:DS exp plots}(f)) is also slightly biased towards the lower anti-diagonal. Due to the scattered nature of the two-photon densities, simple Fourier filtering was not sufficient to filter noise from the experimentally obtained JPDs, and a much more involved cleaning technique was used, as described in appendix \ref{app:bloom}. It is mathematically straightforward to show that the functional form of the excess correlation $\Delta G^{(2)}$ from phase-correlated light is directly related to the correlated phase front structure. Thus, this experiment also serves as a basis for understanding interference in phase-correlated light.

\section{Conclusion and Discussions}
\label{sec:conclusion}
In conclusion, we have explored the full propagation of the biphoton wavefunction in an infinite domain, folded onto a finite region. This has enabled us to experimentally investigate the propagation of the SPDC biphoton state and measure the biphoton densities. We explicitly show how the correlations are transferred continuously from position to momentum in a single experimental setup. We also identify the plane, accurately up-to few millimetres, where the correlation exists only in the phase of the wavefunction. This plane is where the entanglement is known to be completely transferred to the phase. Using classic interference experiments, we have provided direct and conclusive proof of the existence of phase entanglement, and also extracted experimental features previously unexplored in this domain. Since most applications involving spatial entanglement rely on interference effects using these states, we believe that our novel experimental exploration of interference effects can pave the way for applicative experiments in this domain.

\par Most studies utilizing SPDC explore either position correlation or momentum anti- correlation. Consequently, an extensive investigation into phase-correlated states remains lacking. The characteristics of phase-entangled quantum light have not been explored to their potential. This situation is a rather non-intuitive one, with two photons simultaneously in the plane wherein each photon has undetermined phase, but nonetheless, the joint phase of the two photons is deterministic. Therefore, an experimental technique towards the exact location of phase-entangled planes has great potential significance in quantum photonic studies.    A full-state quantum holography of a phase-entangled state, similar to \cite{Devaux2019,Zia2023}, could provide deeper insights, particularly when using a type-II crystal, which facilitates polarization-based holography \cite{Defienne2021}.  Nonetheless, as we have demonstrated, a simple DS experiment is sufficient to highlight the fundamental physics underlying this state. One major application of such a state is akin to ghost imaging, as in \cite{Aspden2016}, where one photon passes through a phase object and is detected by a bucket detector, while the other photon, due to its phase correlation, can be used for holography to retrieve the phase changes induced by the non-locally correlated wavefront. This approach could enable phase imaging with a few photons in Differential  Interference Contrast (DIC) microscopy-type experiments, or aid in estimating the complex phase transmission matrix of disordered media \cite{Gnatiessoro2019,Devaux2023,Courme2023}.
 \par Additionally, SPDC sources have recently been employed in Fourier ptychography \cite{Aidukas2019} and position-momentum correlated imaging \cite{Karimi2024}, where the bi photon's position and momentum information are directly exploited for imaging. It would be of interest to evaluate whether such states could be particularly useful for imaging phase-type targets via ptychography or position-momentum based phase retrievals.
 \par Finally, it is well established that the initial pump phase front is precisely replicated by the SPDC biphoton wavefunction \cite{Riberio1999}. The phase-entangled state enables direct access to this phase front, which could be used to investigate phase changes introduced by phase objects illuminated by the pump beam but analysed using the SPDC beam, in a manner similar to \cite{Hugo2024}. Such an approach could have significant implications in the study of disordered media \cite{Kiran2025}. 

\section*{Acknowledgement}
We acknowledge the Department of Atomic Energy, Government of India, for funding for Project Identification No. RTI4002 under DAE OM No. 1303/1/2020/R\&D-II/DAE/5567, Ministry of Science and Technology, India. 
\section*{Disclosure}
The authors declare that there are no conflicts of interest related to this article.

\section*{Data Availability}
Data and related code for analysis for this procedure is available upon reasonable request.
\appendix
\section{Blooming modelled as double Gaussian and statistical noise correction}
\label{app:bloom}
\begin{figure}[!htbp]
    \centering
    \includegraphics[width=1\linewidth,trim =0.25cm 0.25cm 0cm 0.25cm,clip]{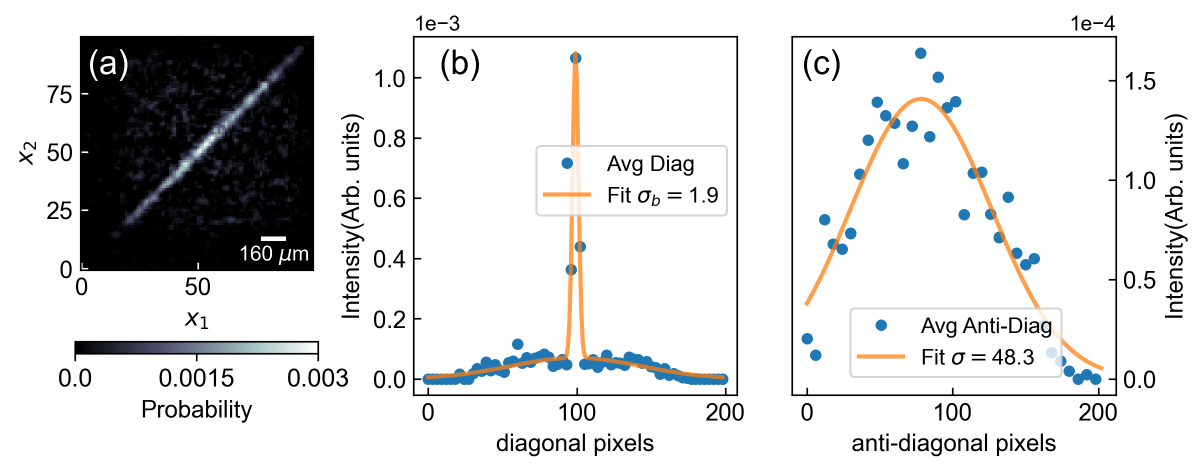}
    \caption{ Figure to depict blooming is DG in nature (a) Noise filtered blooming data obtained from Fig.\ref{fig:filter}(b). (b) Diagonal mean and (c) Anti-diagonal mean of (a)}.
    \label{fig:bloom_app}
\end{figure}
To understand the consequence of blooming on the measured data, we make incident a spatially uncorrelated photon source on the EMCCD, ensuring all parameters remain same as  during data acquisition. This is achieved by using a polarizing beamsplitter in front of the biphoton source and doubling the pump power, thus allowing only one of each pair to reach the EMCCD and yet maintaining the mean photon flux.  Computing the JPD using Eq. \eqref{eqn:hugo} with $3 \times 10^5\text{ frames}$ (Fig. \ref{fig:filter}(b)) gives the form of artificial correlation occurring due to blooming. The 2D JPD for blooming is shown in Fig. \ref{fig:bloom_app} (a). This is the background-cleaned version of \ref{fig:filter}(b). To fit a DG to the bloom, we first create two 1D plots by averaging over the diagonal and anti-diagonal. As seen in Fig. \ref{fig:bloom_app}, the two plots are well-fitted by Gaussians of standard deviation $\sigma_b\approx 1.9$ ($\sigma\approx 48.3$) for the diagonally (anti-diagonally) averaged data. Note that $\sigma$ is found to be the same as the beam size ($\sigma_\text{beam}$). Thus functionally, blooming can be modelled as a DG, labelled $\rho_{\text{bloom}}$, with widths $\sigma_+\to\sigma_\text{beam}$ and $\sigma_-\to\sigma_b$. The $\sigma_{\text{beam}}$  for  each $\bar{z}$ is obtained by fitting a Gaussian to the SPDC beam as it changes with propagation, unlike $\sigma_{b}$ which remains constant. A typical $\rho_{\text{bloom}}$ is shown in Fig~\ref{fig:filter}(c). The appropriately-weighed $\rho_{\text{bloom}}$  is subtracted from Fig~\ref{fig:filter}(a) to obtain filtered data shown in Fig~\ref{fig:filter}(d). Alternately, blooming can also be reduced by spatially separating the two photons, either by polarization separation or non-collinear SPDC.
\par But blooming is not the only type of noise that affects the fidelity of the data, as discussed in sec.~\ref{sec:bloom and background}. Noise from finite sampling—such as statistical fluctuation—and structural noise—such as the background marginal distribution embedded within the original two-photon density—must be removed to accurately estimate the true two-photon distribution. While simple bandpass Fourier filtering suffices for cleaning propagation densities, the scattered nature of two-photon interference densities demands a more nuanced approach to eliminate the embedded marginal structure. To address this, we employed discrete wavelet analysis, which decomposes an image into separate low- and high-frequency components—conventionally referred to as approximation and detail images, respectively. For structural decomposition, we use Daubechies wavelets, known for robust edge detection in noisy environments \cite{Li2025}. To extract the marginal structure, we compute the marginal density as  $\sum_{x_i,y_i}\Gamma_{x_i,y_i;x_j,y_j}$, where $\Gamma$ is the 4D JPD obtained from Eq.~\ref{eqn:hugo} as described in sec.\ref{sec:computing JPD}. It was observed that the marginal density possesses significantly lower spatial frequency than the two-photon interference patterns of interest.

\par Consequently, we perform power spectral analysis on the approximation images of both the JPD and the marginal. Using the marginal’s spatial bandwidth, we apply a high-pass Fourier filter to the JPD’s approximation image to isolate the interference pattern from the marginal background. High-frequency fluctuation noise is directly countered by filtering the detail images using established methods like total variation de-noising \cite{Rudin1992}. Finally, the cleaned approximation and detail images are recombined via inverse wavelet transform to reconstruct the output density. The output density is smoothened by using a Gaussian kernel density estimator with minimal bandwidth. This process yields the final interference patterns, as shown in Fig.\ref{fig:DS exp plots}(a) and Fig.\ref{fig:DS exp plots}(d), using position- and phase-correlated two-photon light, respectively.    
\section{Double Gaussian fits along Diagonals}
\label{app:DG}
\begin{figure*}[!htbp]
    \centering
    \includegraphics[width=0.85\linewidth,trim =0cm 0cm 0cm 0cm,clip]{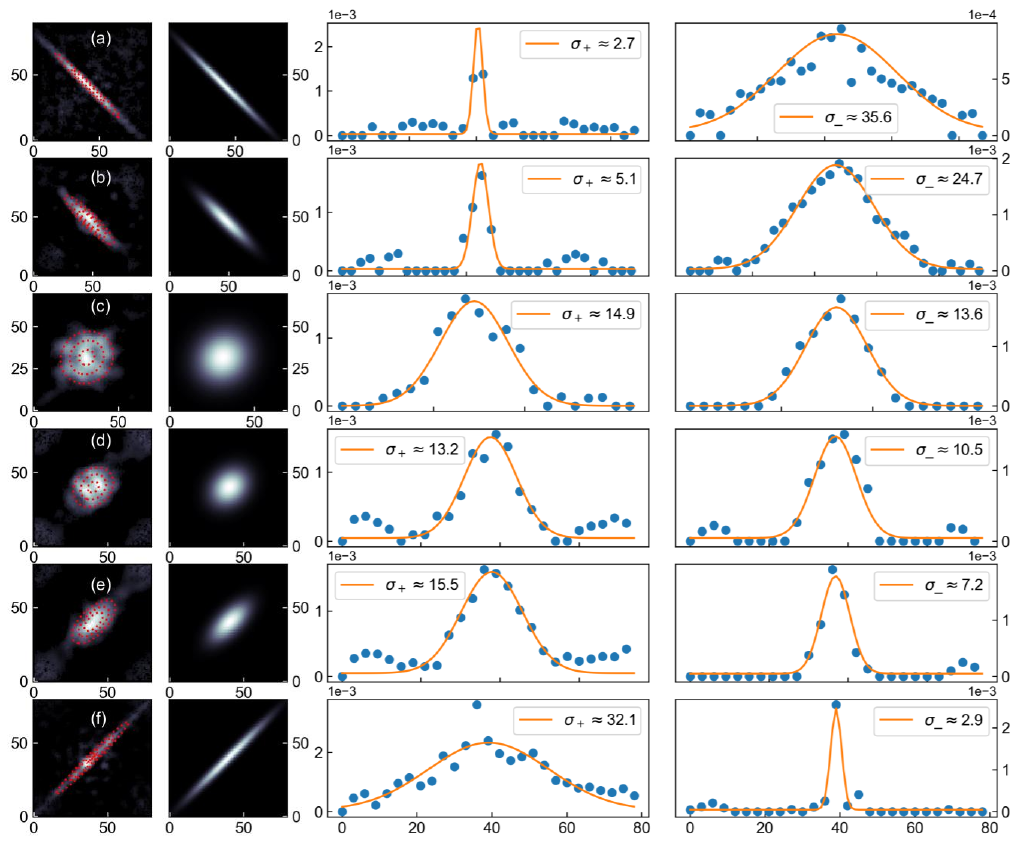}
    \caption{The figure shows the explicit DG fitted on each of the data points of Fig. \ref{fig:jpd}. Each four picture row consists of the DG fitted to the data followed by the slice along the anti-diagonal representing the width $\sigma_+$ and the slice along the diagonal representing $\sigma_-$. The distances are same as Fig. \ref{fig:jpd} (a) $f_3 $ (b) $65$ mm (c) $85$ mm (d) $100$ mm (e) $115$ mm (f) $3f_3$ }
    \label{fig:jpd_app}
\end{figure*}

The DG at various $\bar{z}$ (from Fig.~\ref{fig:jpd}) and its diagonal and anti-diagonal slices are shown in Fig.~\ref{fig:jpd_app}. The DG and its fits (red dotted) are shown in the first column while the numerically calculated JPD for these fit parameters are shown in the second column. It is observed that, as we approach the phase entanglement plane, the computed JPD shown in the first column exhibits spoke patterns and seemingly does not develop smoothly into Gaussians. These patterns arise from limitations in the Fourier filtering process and the subsequent smoothing of data used to remove background statistical noise. However, this structure has minimal impact on the results, as the distributions are fitted based on their overall shape and extent rather than their internal fine structure. One can observe the fidelity of the DG fits, as shown in the third (fourth) column showcasing the anti-diagonal (diagonal) slice, with the orange overlays of the DG fit parameters. Note that only a few data points are shown (blue dots) to avoid clutter.

\end{document}